\documentclass[%
 preprint,
 showpacs,preprintnumbers,
 amsmath,amssymb,
 aps,
 pra,
]{revtex4-1}

\usepackage{graphicx}
\usepackage{dcolumn}
\usepackage{bm}
\usepackage[utf8]{inputenc}
\usepackage[mathlines]{lineno}
\usepackage[english]{babel}

\begin{document}
\preprint{APS/123-QED}

\title{Transport in a chain of asymmetric cavities:\\Effects of the concentration with hard-core interaction.}
\author{Gonzalo Su\'arez}
\author{Miguel Hoyuelos}
\author{H\'ector O. M\'artin}

\affiliation{Departamento de F\'{\i}sica, Facultad de Ciencias Exactas y Naturales, Universidad Nacional de Mar del Plata} 
\affiliation{Instituto de Investigaciones F\'{\i}sicas de Mar del Plata (IFIMAR -- CONICET), De\'an~Funes~3350, 7600~Mar~del~Plata, Argentina}%

\date{\today}

\begin{abstract}
We studied the transport process of overdamped Brownian particles, in a chain of asymmetric cavities, interacting through a hard-core potential. When a force is applied in opposite directions a difference in the drift velocity of the particles inside the cavity can be observed. Previous works on similar systems deal with the low concentration regime, in which the interaction is irrelevant. In this case it was found that large particles show a stronger asymmetry in the drift velocity when a small force is applied, allowing for the separation of different size particles (Reguera \textit{et al}., Phys. Rev. Lett \textbf{108}, 020604, 2012). 
We found that when the interaction between particles is considered, the behavior of the system is substantially different. For example, as concentration is increased, the small particles are the ones that show a stronger asymmetry.  For the case where all the particles in the system are of the same size we took advantage of the particle-vacancy analogy to predict that the left and right currents are almost equal in a region around the concentration $0.5$ despite the asymmetry of the cavity.
\end{abstract}
   
\pacs{05.40.-a, 05.60.-k, 66.10.Cg, 05.10.Ln}  
\maketitle


\section{Introduction}
The diffusion of particles through narrow channels is a phenomenon vastly seen in nature and in artificial devices. In systems such as biological cells, ion channels, nano-porous materials and zeolites the transport of small particles through narrow channels is present. \cite{zwanzig,Brangwynne2009,karger,kosinka,berdakin2013}

In particular, the problem of transport of particles in different kinds of periodic channels has been studied from several points of view. For example, analytic solutions were found to one-dimensional reducible and smoothly corrugated channels, making use of the Ficks-Jacob equation \cite{rubi2010,burada2008,hanggi2009,Ai2006}. Feng-guo Li and Bao-quan Ai \cite{feng2013}, studied the transport of overdamped Brownian particles in channels with curved midline. When this channels are not smoothly corrugated they are called compartmentalized or septate channels. Ghosh \textit{et al.} numerically investigated the transport of particles in symmetric septate channels \cite{ghosh2012,ghosh2011}. Marchesoni \textit{et al.} investigated the flux of particles in an asymmetric septate channel, with reflecting boundary conditions \cite{marchesoni2009}. The dependence of the effective velocity on the period of an external force was analized by Zitserman \textit{et al.} \cite{zitserman}. Dagdug \textit{et al.} \cite{dagdug} studied the particle's mobility under variation of the asymmetry of the cavities.

In this work we have used an asymmetric cavity, with two small exits, as shown in Fig.~\ref{fig:latt}. A similar cavity was used in \cite{ghosh2013} but using self-propelled Janus particles. Here we deal with Brownian particles that interact with each other through a hard-core potential, one of the simplest interactions that can be considered. We focused in the overdamped regime, because of its relevance in most molecular transport systems. 

For small concentration of particles the transport in asymmetric cavities has been extensively studied using the Langevin equation (see \cite{Reguera2012} and references cited therein). The entropic effects become relevant generating an entropic potential which is responsible for a peculiar behavior. If a force of the same magnitude is applied to the right or to the left, the magnitude of the resulting drift velocities to the right or to the left are different. In these cases the effects of interactions are assumed irrelevant due to small concentrations. The aim of the present work is to extend the study of transport of particles for all concentrations taking into account the hard core interaction. We studied the effect of increasing the concentration and changing the size of the particles.  If the concentration is high enough, it can be expected that the interaction between particles modifies the flux asymmetry. We found that the flux asymmetry is not only modified, but also inverted. 

We used a simplified model to carry out simulations using the Monte Carlo method. Even though the cavity chosen is arbitrary, the arguments and conclusions that we have found are general and independent of specific details of the shape of the cavity. Two different sizes of particles were used to analyze how this feature modifies the transport process. 

This work is organized as follows. In Section \ref{sec:model} the details of the model are explained. In Section \ref{sec:results} we show the results obtained and in Section \ref{sec:analogy} we discuss, using simple arguments, how a particle-vacancy analogy can be used to describe some aspects of the dynamics of the system. Some final remarks are exposed in Section \ref{sec:conclusion}.

\section{Model}
\label{sec:model}

\begin{figure}
 \includegraphics[width=0.9\linewidth]{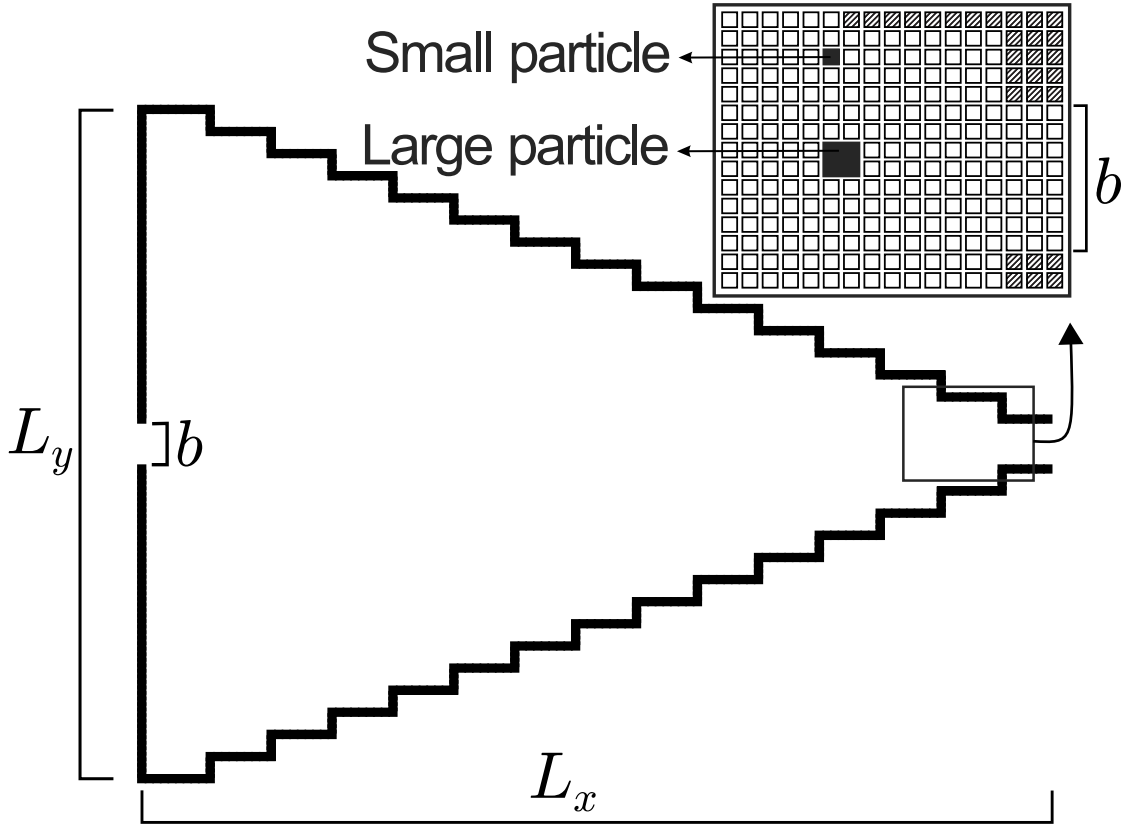}
 \caption{Lattice structure. \\ $b=8a, L_x=120a, L_y=128a$ }
 \label{fig:latt}
\end{figure}

To perform the simulations,  we used the standard Monte Carlo method with a square lattice of size $L_x \times L_y$, with a lattice spacing $a=1$. A triangular area, where the particles can diffuse, has been delimited inside, with two  exits of size $b$ on the sides (Fig. \ref{fig:latt}). Periodic boundary conditions were set on the exits, to simulate a chain of identical cells repeated along the $x$ axis.

Two kinds of particles where used. The small ones that occupy one site of the lattice, and the large ones that occupy four sites (a square of two sites per side). The particles can diffuse freely as long as they do not encounter any other particle or a wall. 

When there are no forces applied, all the particles have the same jumping rate in all directions. The jumps have a length of $a=1$. If there is a force, $\delta$, acting along the $x$ axis, the jumping rate will be higher in the direction of the force. Thus, if the force is applied to the right, the jump rates in different directions are
\begin{equation}
  \begin{split}
    p_\uparrow=p_\downarrow=p_\leftarrow=p\\
    p_\rightarrow=p(1+\delta).
  \end{split}
  \label{eq:rate1}
\end{equation}

Likewise, if the force is applied to the left,
\begin{equation}
  \begin{split}
    p_\uparrow=p_\downarrow=p_\rightarrow=p\\
    p_\leftarrow=p(1+\delta).
  \end{split}
  \label{eq:rate2}
\end{equation}

If a force to the right (left) is applied in the steady state regime the modulus of the mean velocity in the $x$ direction $v^{+\delta}$ ($v^{-\delta}$, respectively) is obtained. The difference $\Delta v=v^{+\delta}-v^{-\delta}$ was used as a measure of the asymmetric effect of the cavity. If it were symmetric, then the difference between $v^{+\delta}$ and $v^{-\delta}$ would be zero. 

We can relate the dimensionless external force $\delta$ used in the model with the  force $F$ acting on the particles. The relation is $(1+\delta)=\exp(Fa/kT)$, where $k$ is the Boltzmann constant. Considering $\delta \ll 1$, we have $\delta \approx Fa/kT$. Nevertheless, we will also consider the case $\delta$ of order 1, in which $\delta$ still represents the force although it is not directly proportional to it.

We consider that small particles have a diffusion coefficient $D=pa^2$ and a force $\delta$ is applied to them in the $x$ direction. The diffusion coefficient for large particles is set to $D'=D/2$ and the force to $\delta'=2\delta$. With this values, in a free space (i.e., in a full square lattice without any restrictions), both kinds of particles have the same mean velocity. For example assuming that the forces are applied to the right
\begin{equation}
 v'_x=p_x'\delta'a=\frac{p_x}{2}2\delta a=p_x\delta a=v_x
\end{equation}

This is the reason why we use the above mentioned values of $D'$ and $\delta'$. 

We analyze stationary situations when a constant force, to the right or to the left, is applied and particles drift in an array of asymmetric cavities, as the one shown in Fig. \ref{fig:latt}. With this constraint, the effective velocity in the steady state might be much smaller than $p_x a \delta$ due to collisions with the cavity walls. A linear relation between effective velocity and force holds for small values of $\delta$. In this linear regime, there is no difference between $v^{+\delta}$ and $v^{-\delta}$, and $\Delta v = 0$ \cite{zitserman}. We are interested in intermediate values of $\delta$, for which the linear regime does not hold and one expects a difference $\Delta v \neq 0$ due to the asymmetry of the cavities and entropic effects \cite{Reguera2012}, at least for small concentrations. More specifically, for the cavity shown in Fig. \ref{fig:latt}, $\Delta v$ for large particles would be greater than $\Delta v$ for small particles, in a range of intermediate values of $\delta$. The extension to higher concentrations will be discussed in the next section.

\section{Results}
\label{sec:results}

In the simulation, we define the concentration $c$ as the mean number of particles in the cavity in the steady state regime divided by $S$ for small particles, and divided by $S/4$ for large particles, where $S$ is the total number of sites in the cavity. We use a value of the jump rate $p$ equal to 1. From now on we denote by $c=0$ the case in which there is only one particle in the cavity. The results of the mean velocity in the horizontal direction were obtained analyzing the slope in a plot of mean displacement against time. The slope is typically measured in the time range $4000 < t < 10000$ in order to avoid the transient and to evaluate the velocity in the steady state. The initial condition is random: particles are homogeneously distributed in the cell, with probability $c$.

\begin{figure}
 \includegraphics[width=0.9\linewidth]{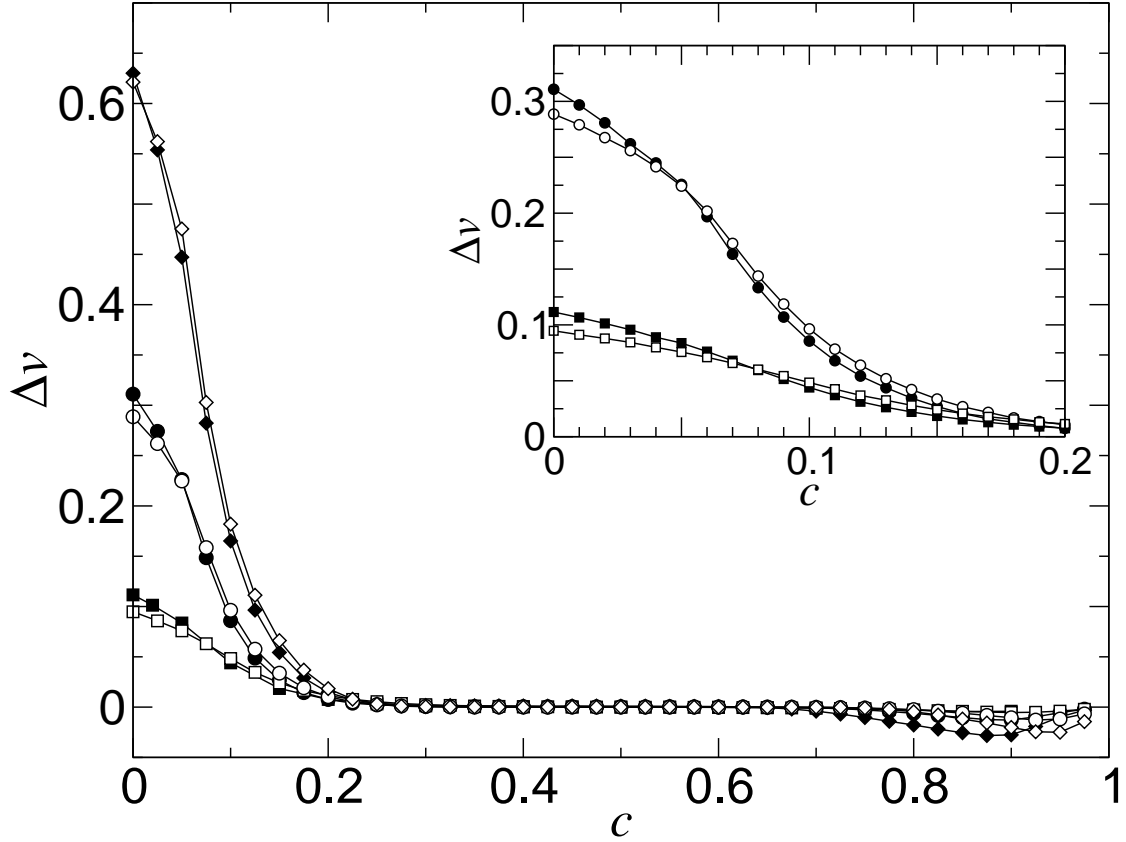}
 \caption{$\Delta v$ as a function of $c$ for $\delta=0.2$ ($\Box$), $\delta=0.5$ ({\large$\circ$}) and $\delta=1$ ($\Diamond$). Filled symbols for large particles and empty symbols for small ones. The inset shows an amplified region in $0\leq c\leq 0.2$. Trajectories (number of samples times number of particles in each sample): $700\,000$.}
 \label{fig:velocity1}
\end{figure}

It is interesting to analyze how  $\Delta v$ changes as the concentration increases, and the interaction becomes more relevant. See Fig. \ref{fig:velocity1}.

For $c=0$ the results are in qualitative agreement with Ref.\cite{Reguera2012}. That is, firstly, for a not too small value of the force, $\Delta v>0$. And secondly, $\Delta v$ for large particles ($\Delta v_\mathrm{large}$, filled symbols in Fig. \ref{fig:velocity1}) is greater than $\Delta v$ for small particles ($\Delta v_\mathrm{small}$, empty symbols). However, when the concentration gets higher, both $\Delta v$ decrease. Even more, for large enough $c$, $\Delta v$ becomes negative. This means that, even when the cavity is designed to favor the transport to the right, the hard-core interaction makes the particles move faster, in average, to the left.

\begin{figure}
 \includegraphics[width=0.9\linewidth]{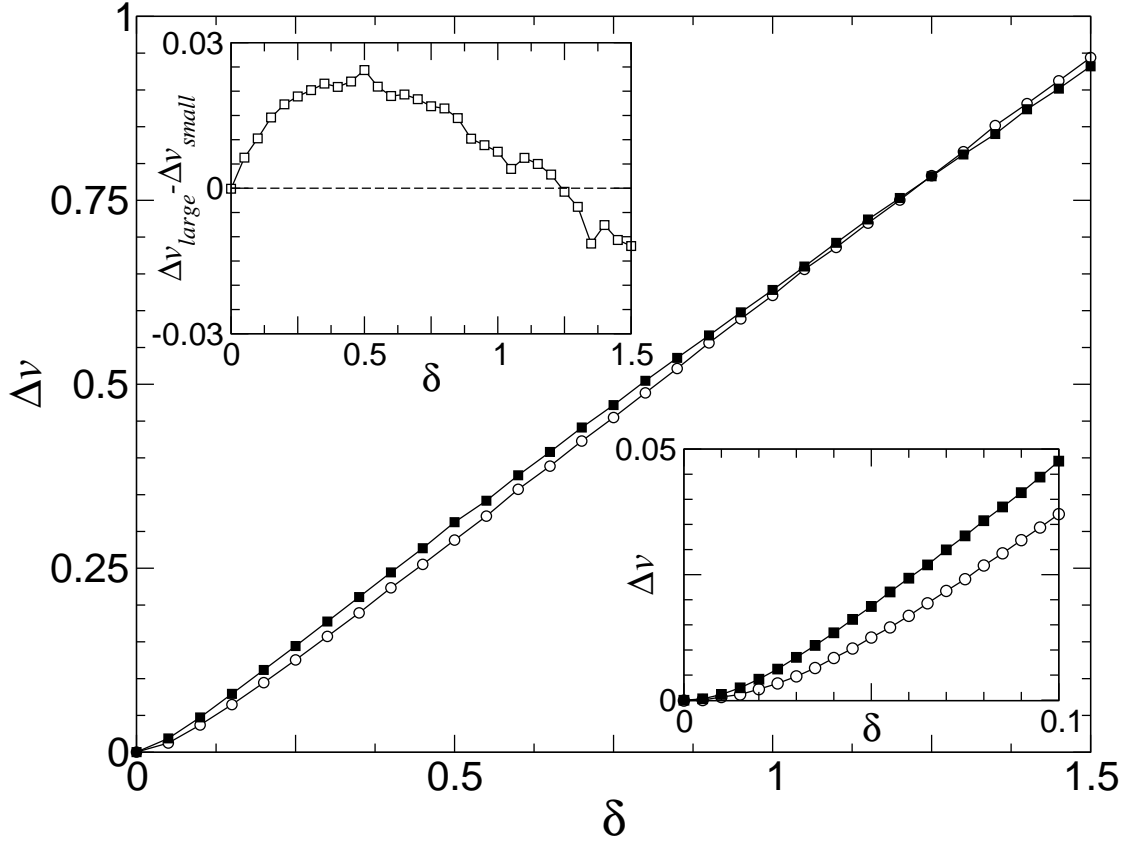}
 \caption{$\Delta v_\mathrm{large}(\blacksquare)$ and $\Delta v_\mathrm{small}$ ({\large$\circ$}) as a function of $\delta$ and $c=0$. Upper inset: $\Delta v_\mathrm{large}-\Delta v_\mathrm{small}$. Lower inset: zoom of $\Delta v$ against $\delta$ for small values of $\delta$. Number of samples: $30\,000$.}
 \label{fig:velocity3}
\end{figure}

Let us now compare $\Delta v_\mathrm{small}$ with $\Delta v_\mathrm{large}$. As expected, for $c=0$ (and not too large values of $\delta$) $\Delta v_\mathrm{small} < \Delta v_\mathrm{large}$. For other concentrations, it can be noticed that there is a value of $c$ for which both curves intersect (see the inset of Fig. \ref{fig:velocity1}), and the relation is inverted.

To determine the value of $\delta$ that exhibits the largest entropic effect, we compared $\Delta v_\mathrm{small}$ and $\Delta v_\mathrm{large}$ for $c=0$.
The results are shown in Fig. \ref{fig:velocity3}. For small values of $\delta$, $\Delta v_\mathrm{large}$ is higher than $\Delta v_\mathrm{small}$, meaning that the entropic effect is bigger for larger particles when the concentration is small. Nevertheless, when the force is strong enough $\Delta v_\mathrm{small} > \Delta v_\mathrm{large}$. See the upper inset of Fig. \ref{fig:velocity3}. The lower inset in Fig. \ref{fig:velocity3} shows that the slope of $\Delta v$ against $\delta$ tends to zero when the force tends to zero, indicating a linear relation between velocity and force for small values of $\delta$ \cite{zitserman}.

\section{Particle-vacancy analogy}
\label{sec:analogy}

We begin by defining the average number of particles that cross the whole cavity per unit of time when a force $\delta$ is applied. Independently of the cavity geometry, we assume that it has a characteristic length $L_x$, and that a particle needs an average time $\Delta t$ to cross the whole cavity. Then, the mean velocity inside the cavity is $v^{+\delta}=L_x/\Delta t$. After a time $\Delta t$, $c S_{\mathrm{eff}}$ particles (where $S_\mathrm{eff}=S$ for small particles, and $S_\mathrm{eff}=S/4$ for large ones) cross the cavity. The mean number of particles per unit time that leave the cavity is $N^{+\delta}_c = c S_\mathrm{eff}/\Delta t = c v^{+\delta}_c S_\mathrm{eff}/L_x$ (subindex $c$ is used to indicate the concentration value for which a given quantity is evaluated).  Defining the average current as $I^{+\delta}_c= N^{+\delta}_c/(\frac{S_\mathrm{eff}}{L_x})$, we have
\begin{equation}
  I^{+\delta}_c= c v^{+\delta}_c.
  \label{eq:particles_intensity}
\end{equation}

In the following, only small particles are considered and subscripts `large' and `small' in $\Delta v$ are omitted.

Our system has a frequently found symmetry: holes behave as particles moving in the opposite direction (see, e.g., \cite{dierl2013,torrez2013}). Making use of this symmetry it is possible to find an easier way to describe the physics of the problem.  We can think on the vacancies as particles moving in the opposite direction, bounded to a force of magnitude $-\delta$, in a system with concentration $1-c$. Because of that, it is possible to write the average current of vacancies,
\begin{equation}
  I^{-\delta}_{1-c}=(1-c)v^{-\delta}_{1-c}.
  \label{eq:vacancy_intensity}
\end{equation}

It is straightforward to notice that the average current of the particles has to be equal to the one of the vacancies. This is a consequence of having the number of particles fixed. Finally, matching equations \ref{eq:particles_intensity} and \ref{eq:vacancy_intensity}, we get
\begin{equation}
  cv^{+\delta}_c=(1-c)v^{-\delta}_{1-c}.
  \label{eq:cv}
\end{equation}

\begin{figure}
 \includegraphics[width=0.9\linewidth]{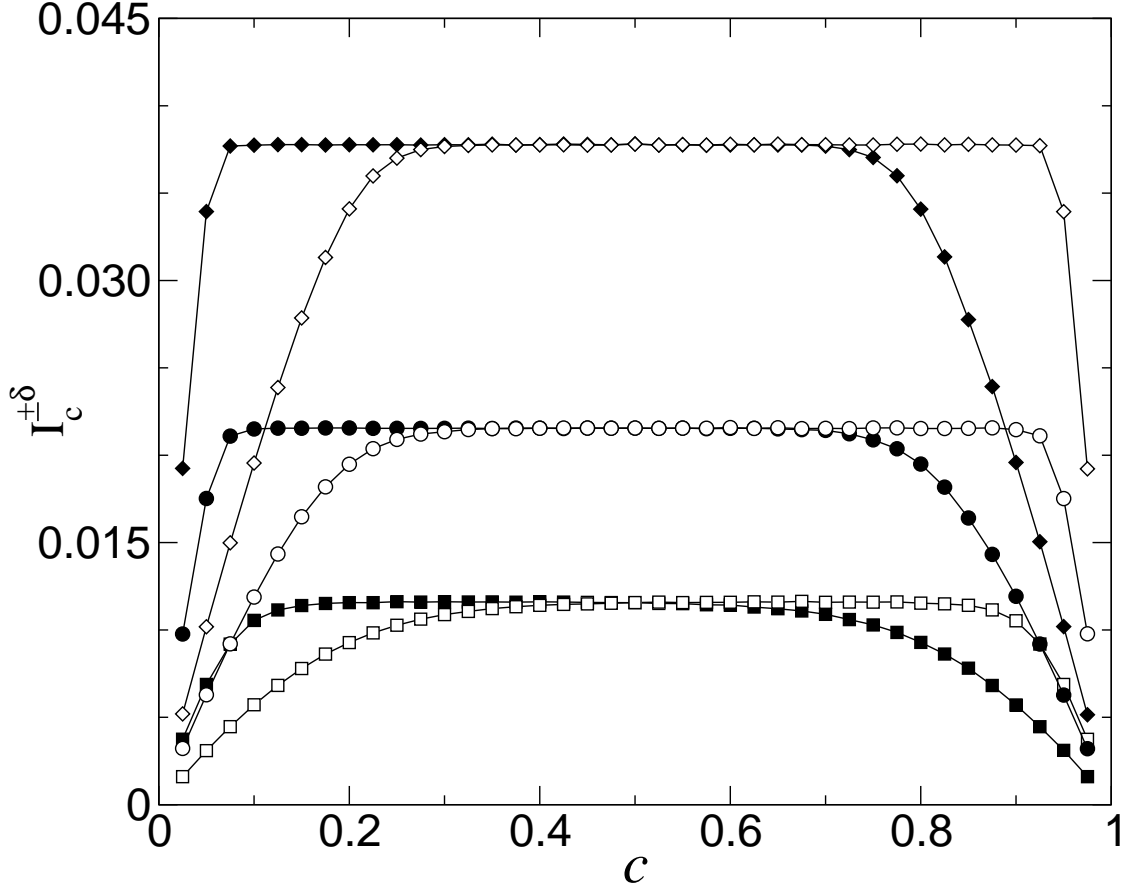}
 \caption{ $I_c^{\pm\delta}$ as a function of $c$ for $\delta=0.2$ ($\Box$), $\delta=0.5$ ({\large$\circ$}) and $\delta=1$ ($\Diamond$). Filled symbols: force to the right. Empty symbols: force to the left, only small particles. Trajectories (number of samples times number of particles in each sample): $700\,000$.}
 \label{fig:intensity1}
\end{figure}

\begin{figure}
 \includegraphics[width=0.9\linewidth]{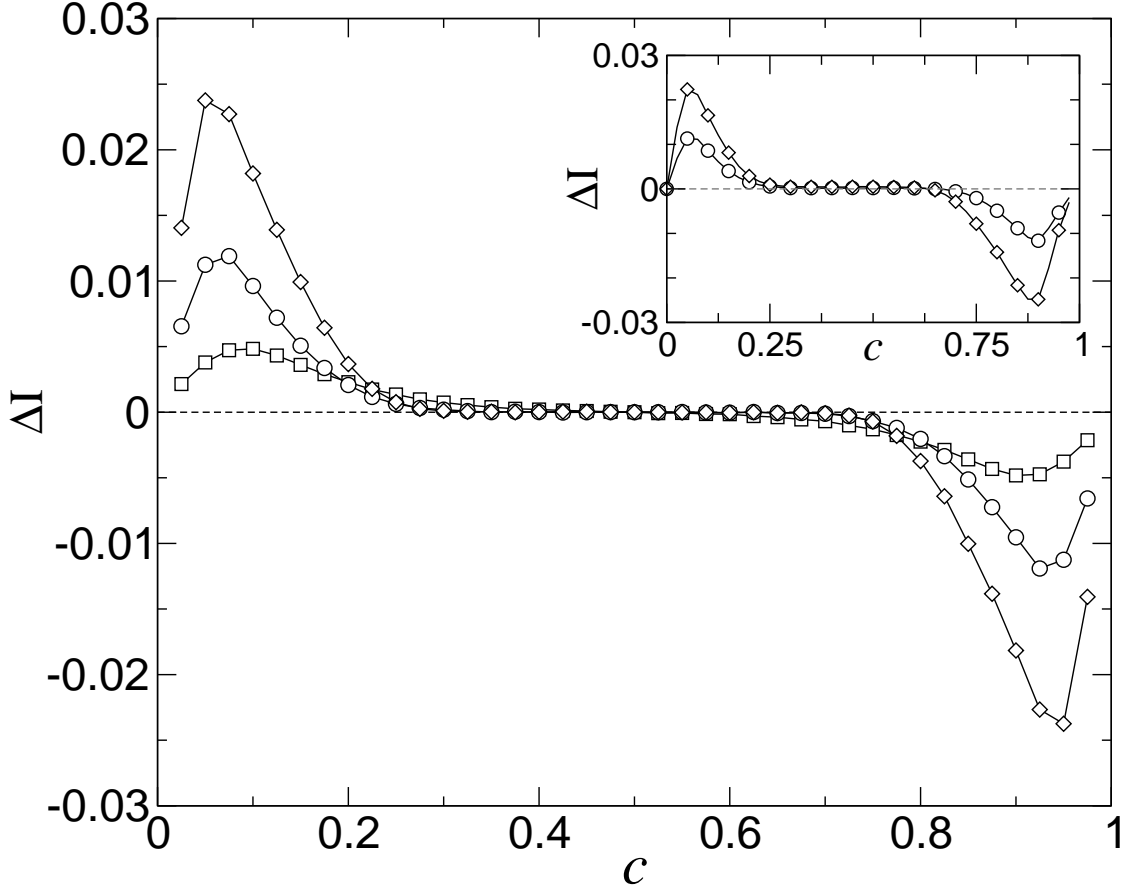}
 \caption{$\Delta I =c\Delta v$ as a function of $c$ for $\delta=0.2$ ($\Box$), $\delta=0.5$ ({\large$\circ$}) and $\delta=1$ ($\Diamond$), only small particles.  Note that the same data is shown in Fig. \ref{fig:velocity1}, but now the symmetry with respect to $c=0.5$ appears because we plot $c\Delta v$ against $c$ instead of $\Delta v$ against $c$. The inset shows $\Delta I$ versus $c$ for large particles only. $\delta=0.5$ ({\large$\circ$}) and $\delta=1$ ($\Diamond$). Trajectories (number of samples times number of particles in each sample): $700\,000$.}
 \label{fig:intensity2}
\end{figure}

The relation (\ref{eq:cv}) is in agreement with what have found with numerical simulations. It explains the symmetry, between force to the right and to the left, of $I_c^{\pm\delta}$ against $c$ shown in Fig. \ref{fig:intensity1} and of $\Delta I$ in Fig. \ref{fig:intensity2}, respect to $c=0.5$. Even more, if $c=0.5$ then $v^{+\delta}_{0.5}=v^{-\delta}_{0.5}$, independently of the magnitude of the force and the shape of the cavity. However, the actual form of $v_c^\delta$ depends on the shape of the cavity. In particular, if the cell is symmetric then $\Delta v = 0$ $\forall c \in (0;1)$. Also, from Fig. \ref{fig:intensity2} we can see that there is an optimal value of $c$ for each value of $\delta$ that maximizes the difference $\Delta I = c \Delta v$.

\begin{figure}
    \begin{minipage}[b]{0.45\linewidth}
      a)\includegraphics[width=0.92\textwidth]{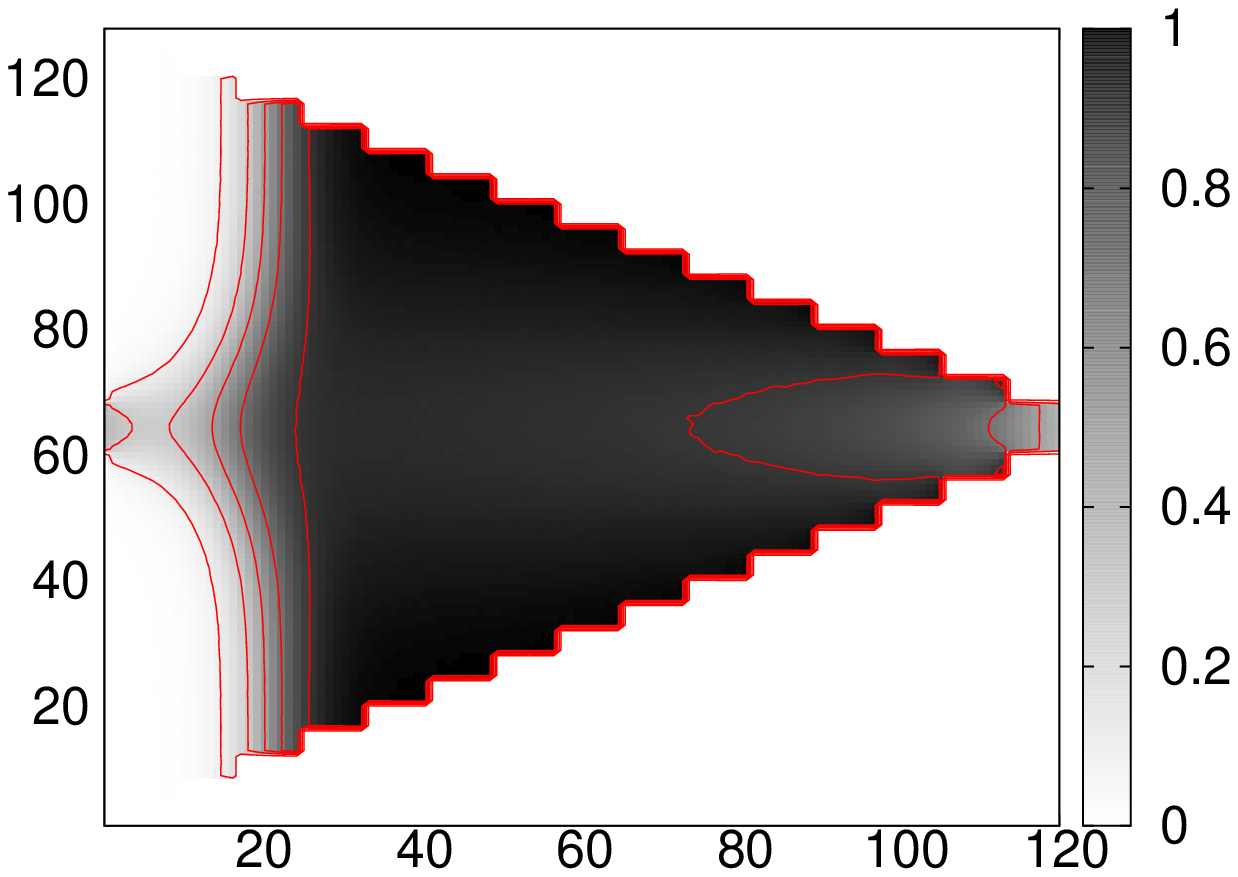}
    \end{minipage}
  \hspace{0.5cm}
    \begin{minipage}[b]{0.45\linewidth}
      b)\includegraphics[width=0.92\textwidth]{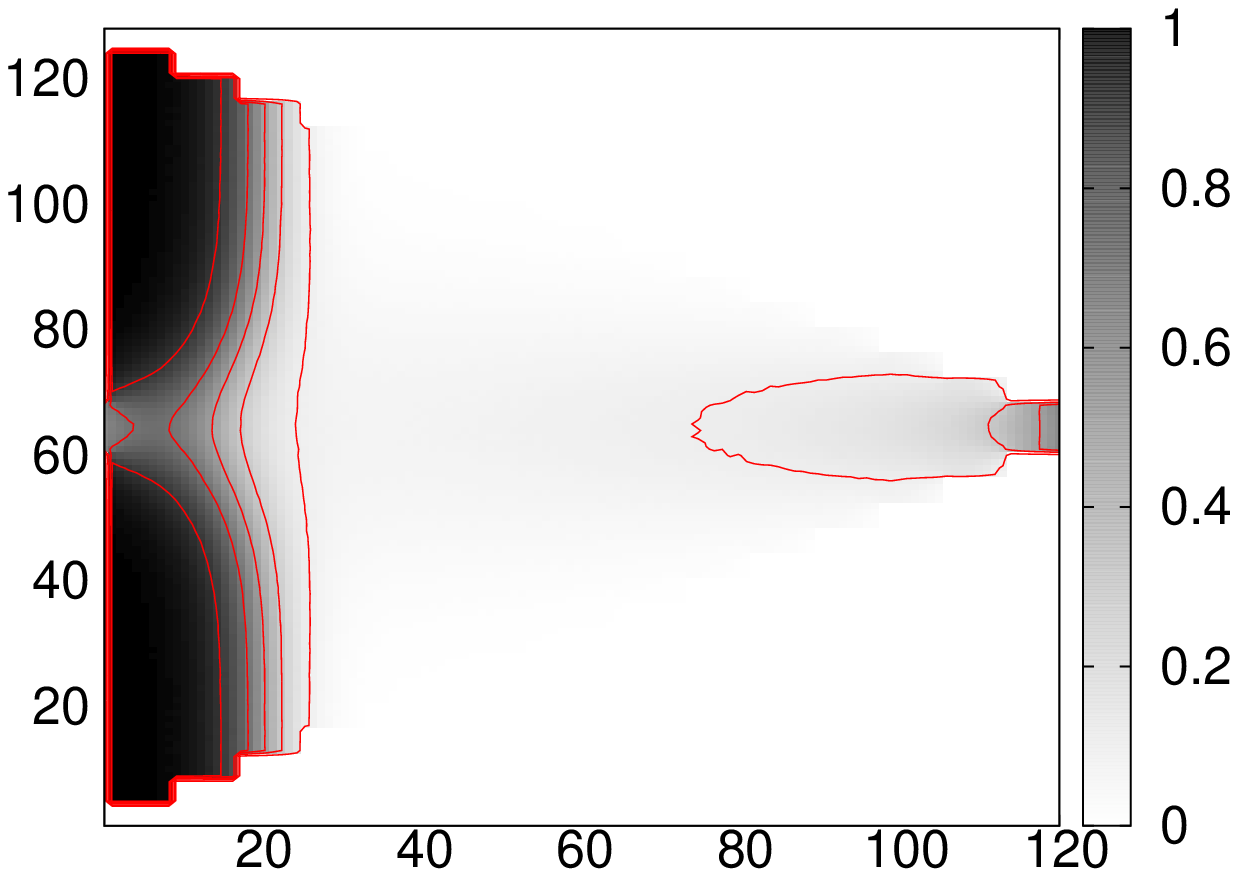}
    \end{minipage}
  \caption{(color online) Average particle density or site occupation frequency $P^{\delta,c}_p$ against position in the cavity. Small particles. Level curves represent equal density. (a) $\delta=+0.5$, $c=0.7$; (b) $\delta=-0.5$, $c=0.3$. Black: occupied site; white: empty site. Time average of $10^5$ snapshots in the range $10^2 < t < 10^8$.}
  \label{fig:0703}
\end{figure}

\begin{figure}
      \begin{minipage}[b]{0.45\linewidth}
        a)\includegraphics[width=0.92\textwidth]{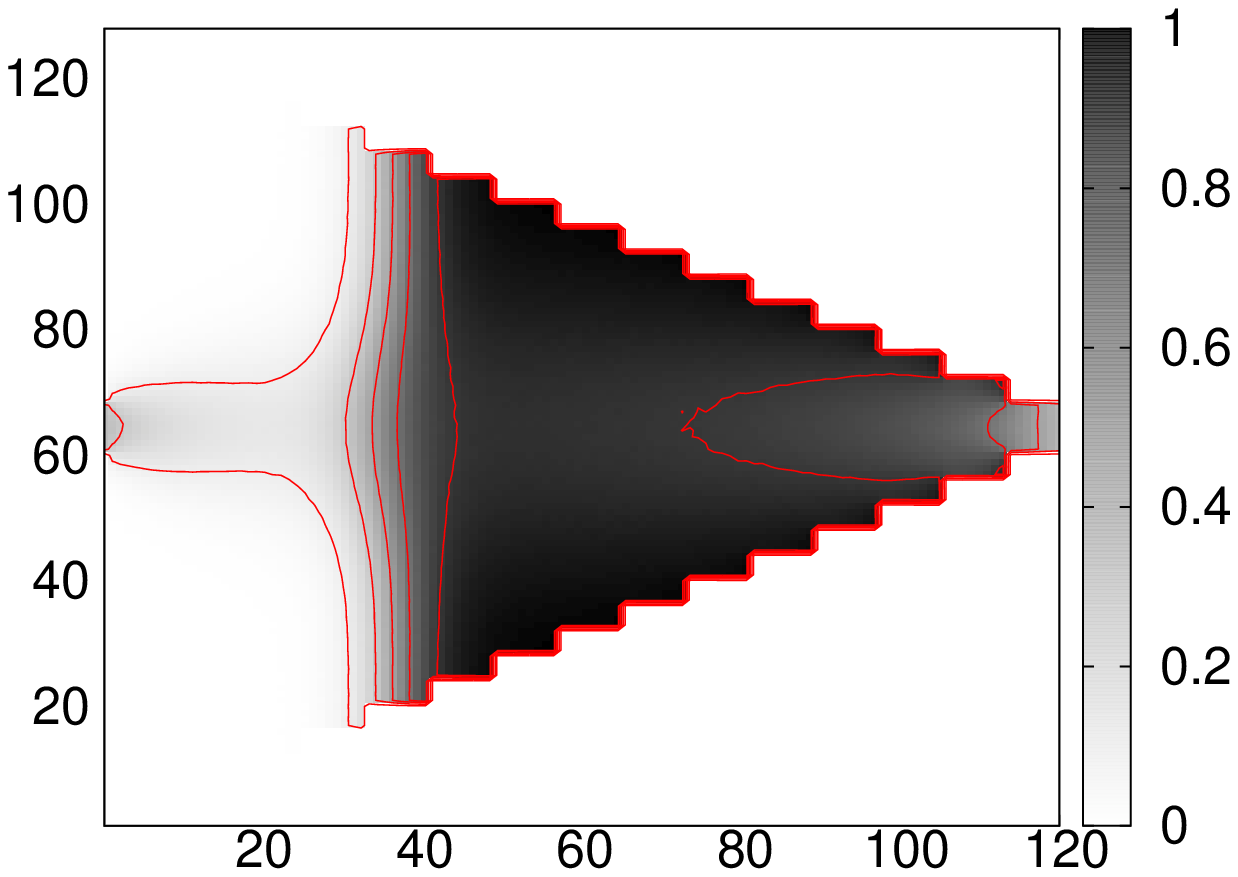}
      \end{minipage}
  \hspace{0.5cm}
      \begin{minipage}[b]{0.45\linewidth}
        b)\includegraphics[width=0.92\textwidth]{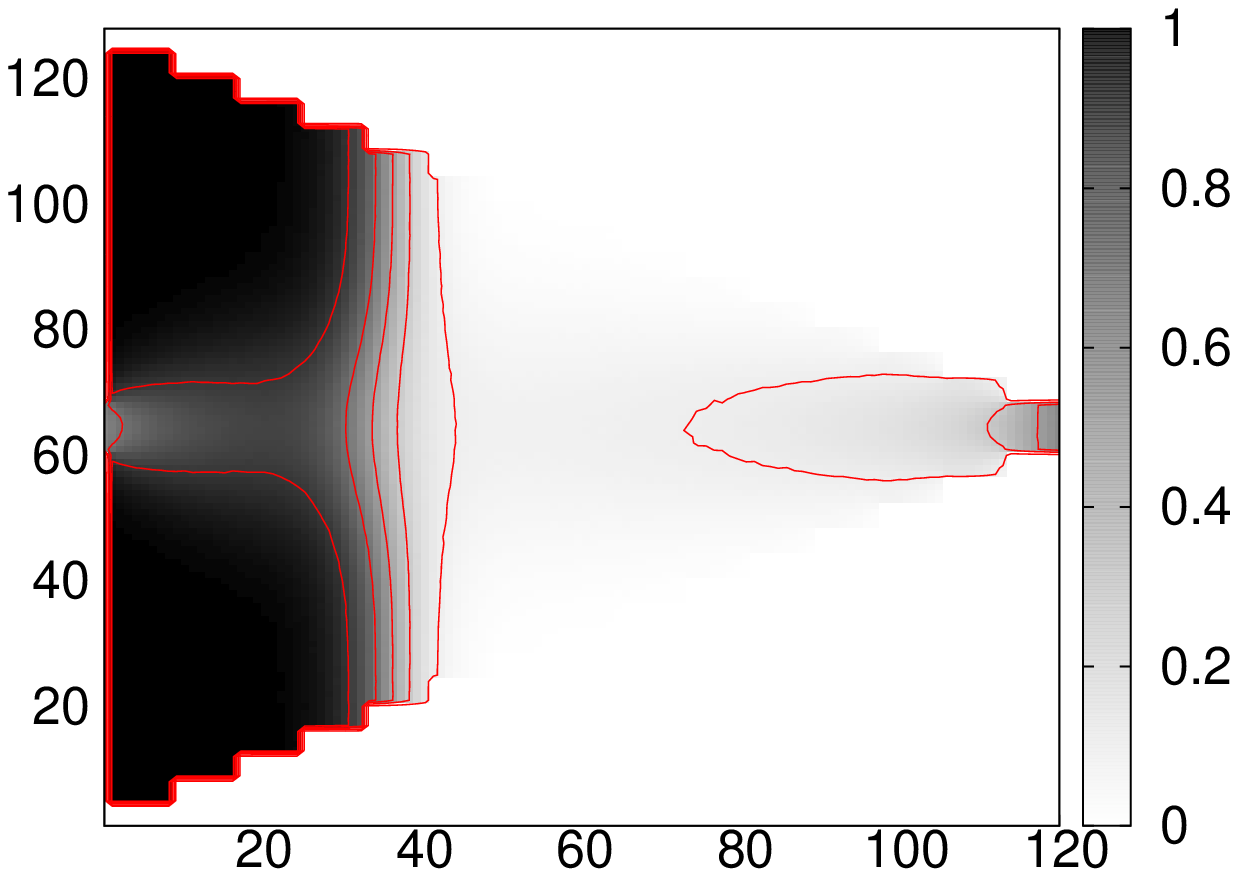}
      \end{minipage}
  \caption{(color online) Average density or site occupation frequency. Small particles. Level curves represent equal density. (a) $\delta=+0.5$, $c=0.5$; (b) $\delta=-0.5$, $c=0.5$. Black: occupied site; white: empty site. Time average of $10^5$ snapshots in the range $10^2 < t < 10^8$.}
  \label{fig:0505}
\end{figure}

Another feature that can be noticed from Fig. \ref{fig:intensity1} is that $I_c^{\pm \delta}$ is constant in a significant region of values of the concentration around $c=0.5$. This indicates that increasing the concentration a little does not modify the current of particles coming out of the cell per unit time. It can be seen that the extra particles increase the size of the clog but do not modify the situation close to the exit hole. This situation is illustrated in Fig. \ref{fig:0703}(b) ($c=0.3$)  and Fig. \ref{fig:0505}(b) ($c=0.5$), where the average density is shown in a stationary situation against position in the cavity. Increasing the concentration $c$ makes larger the black zone of occupied sites in a region far from the exit, that does not modify the current $I_c^\delta$ (see Fig. \ref{fig:intensity1}, empty circles, for values of $c$ close to 0.5). 

The average current is a global magnitude that involves the average of all sites. Now we are going to focus on what happens locally using the site occupation frequency.

Because each site can be found in two states (occupied or not), we have that 
\begin{equation}
    P^{+\delta,c}_p(i,j) + P^{+\delta,c}_h(i,j)=1,
\end{equation}
where $P^{+\delta,c}_p(i,j)$ is the probability of finding the site $(i,j)$ occupied, and $P^{+\delta,c}_h(i,j)$ is the probability of finding the same site vacant. 
Now, knowing that holes are equivalent to particles moving in the opposite direction, we can write $P^{+\delta,c}_h(i,j) = P^{-\delta,1-c}_p(i,j)$, and
\begin{equation}
   P^{+\delta,c}_p(i,j) + P^{-\delta,1-c}_p(i,j)=1
   \label{eq:sum}
\end{equation}

Equation (\ref{eq:sum}) implies a relation between two cases with different values of the force and concentration. The numerical results of Figures \ref{fig:0703} and \ref{fig:0505} confirm this relation. Case (b) in each figure corresponds to the change $\delta \rightarrow -\delta$ and $c \rightarrow 1 - c$ with respect to case (a).  The plot of the resulting average density in (a) is the negative picture of (b).  Figures \ref{fig:0703}(b) and \ref{fig:0505}(b) show that, for a negative force, particles accumulate to the left mainly in the upper and lower corners of the triangular cavity producing a more symmetric effective shape.  This effect is responsible of the symmetric behavior that we observe in Fig. \ref{fig:intensity2}, where $\Delta I =0$ for intermediate values of $c$.

Let us stress that the particle-vacancy analogy is strictly valid only for small particles due to the particular way to simulate the diffusion processes where the jump length is equal to the linear size of the particle. For large particles this analogy is no longer valid. But even for this case we obtained similar results. For example, in the inset of Fig. \ref{fig:intensity2} we show $\Delta I$ as a function of $c$ for large particles. Although the symmetry respect to $c=0.5$ does not hold exactly, the curves are quasi symmetrical and $\Delta I$ is almost zero around $c=0.5$. In general we expect that the results of the present section are at least qualitatively right for the transport of overdamped Brownian particles, independently of the specific rules of the model definition.

\section{Conclusion}
\label{sec:conclusion}

In this work we investigated the transport of overdamped Brownian particles in a chain of asymmetric cavities, interacting through a hard-core potential. Although there is an entropic effect that makes particles of different sizes diffuse differently, we found that the hard-core interaction diminishes this effect. We compared the dynamic of small and large particles and analyzed the difference, $\Delta v$, between the steady state velocities that are obtained when a force to the left or to the right is applied.  This difference is a measure of the asymmetric behavior of the system.  We qualitatively reproduced previous results for $c=0$ and not too small values of the force, for which the velocity difference for small particles is smaller than the one for large particles, $\Delta v_\mathrm{small} < \Delta v_\mathrm{large}$ (for $\delta \lesssim 1.25$).  As $c$ is increased, the difference not only decreased, but it was inverted at some point. A similar effect was observed when we kept $c=0$ and increased the force $\delta$.

For the case when there is only one kind of particle present in the system, we took advantage of the particle-vacancy analogy to show that there should be a symmetry in the behavior of the average current when the force is inverted and the concentration $c$ is replaced by $1-c$. This symmetry is strictly valid only for small particles and it is due to the specific rules of the model definition. Even with this limitation, the results obtained are qualitatively right independently of the model used.
That is, the difference, $\Delta I$, of the current when the force is applied to the left or to the right has a maximum for a value of $c$ close to 0 and a minimum for $c$ close to 1 (see Fig. \ref{fig:intensity2}).  These values of the concentration maximize the asymmetric effect of the cavity. For $c$ close to $0.5$, there is a range of values of the concentration for which $\Delta I \simeq 0$, this means that, in this case, the asymmetry of the cavity has no effect on the particle current.

\begin{acknowledgments}
This work was partially supported by Consejo Nacional de Investigaciones Cient\'{\i}ficas y T\'{e}cnicas (CONICET, Argentina, PIP 0041 2010-2012).
\end{acknowledgments}


\end{document}